\documentclass[aps,draft]{revtex4}
\usepackage[dvips]{graphicx}
\begin{document}

\title{Excitation of Quantized Longitudinal Electric Waves in a Degenerate Fermi Gas }
\author{ Levan N.Tsintsadze }
\thanks{Also at Department of Plasma Physics, E.Andronikashvii Institute of Physics, Tbilisi,
Georgia}
\affiliation{Graduate School of Science, Hiroshima University, Higashi-Hiroshima, Japan}

\date{\today}

\begin{abstract}

The system of electron beam - degenerate Fermi gas in a magnetic field is investigated.
Instabilities of the quantized longitudinal electric waves are studied by a newly derived dispersion equation. Novel branches of longitudinal waves are found, which have no analogies without the Landau quantization. Growth rates of these new modes are obtained. The excitation of the zero sound by an electron beam is discussed and found that the quantization of the energy of electrons imposes a new condition. Furthermore, the excitation of Bogolyubov's type of spectrum by a strong electric field is considered.

\end{abstract}


\maketitle
Quantum plasma is becoming of increasing current interest motivated by its potential application in modern technology, e.g. metallic and semiconductor nanostructures - such as metallic nanoparticles, metal clusters, thin metal films, spintronics, nanotubes, quantum well and quantum dots, nano-plasmonic devices, quantum x-ray free-electron lasers, etc. Moreover, quantum plasmas are common in planetary interiors, in compact astrophysical objects (e.g., the interior of white dwarf stars, magnetospheres of neutron stars and magnetars, etc.), as well as in the next generation intense laser-solid density plasma experiments. The quantum plasma may arise when a pellet of hydrogen is compressed to many times the solid density in the fast ignition scenario for inertial confinement fusion.

The field of quantum plasma physics is evolving. In the past the properties of linear electron oscillations in a dense Fermi plasma have been studied in Refs.\cite{gol}-\cite{bp}. Despite extensive theoretical efforts (for recent review see \cite{pad}) since then, there are many questions and issues which one has to address. Recently, we have derived a new type of quantum kinetic equations of the Fermi particles of various species, and a general
set of fluids equations describing the quantum plasma was obtained \cite{tsin}. This novel kinetic equation for the Fermi quantum plasma was used in Ref.\cite{tsin} to study the propagation of small longitudinal perturbations in an electron-ion collisionless plasmas, deriving a quantum dispersion equation. The dispersion properties of electrostatic oscillations in quantum plasmas have been discussed very recently in Refs.\cite{tsin10}, \cite{ell}. The effects of the quantization of the orbital motion of electrons and the spin of electrons on the propagation of longitudinal waves in the quantum plasma have been also reported \cite{ltsin}. As is well known, the strong magnetic field in the Fermion gas leads to two magnetic effects \cite{land}. Namely, they are the Pauli paramagnetism due to the spin of electrons and the Landau diamagnetism due to the quantization of the orbital motion of electrons. In the previous work \cite{ltsin}, we have derived a novel dispersion relation of the longitudinal wave propagating along a magnetic field, which exhibits the strong dependence on the magnetic field in radical contrast to the classical case. The magnetic field-aligned electron plasma waves have been considered also in the remote past by Kelly \cite{kel}. But his consideration is not self-consistent, because he assumed $h=0$ (where  h is the Planck constant) and from the Wigner equation  he got the classical Vlasov  kinetic equation, however he took into account the Landau quantization in the Wigner's equilibrium distribution function. Here arises the question how the Wigner distribution function can satisfy the Vlasov kinetic equation. Moreover, Kelly could not explicitly obtain the dispersion equation. He also didn't take into account the influence of de Broglie waves in his investigation.

The affect of strong or superstrong magnetic field on the thermodynamic properties of medium and the propagation of proper waves is an important issue in supernovae and neutron stars, the convective zone of the sun, the early prestellar period of the evolution of the universe, as well as in the laboratory plasmas (the contemporary problems of laser-matter interaction). Based on the astrophysical data, the surface magnetic field of a neutron star is $H\sim 10^{11}-10^{13}G$, and the internal field can reach $H\sim 10^{15}G$ or
even higher \cite{lan}-\cite{lip}. It was shown in Ref.\cite{bis} that the presence of rotation of stars may increase the magnetic field by an additional factor of $10^3-10^4$. In such strong magnetic fields, it is expected that the thermodynamic properties and wave dynamics in degenerate plasmas would be quite different governed by the quantum effects, as discussed in Refs.\cite{ltsin}, \cite{kel}. This is true when the characteristic energy of electron on a Landau level reaches the nonrelativistic limit of the electron chemical potential $\mu=\varepsilon_F=\frac{\hbar\mid e\mid H}{2m_ec},$ i.e.
$H=H_S\frac{v_F^2}{c^2}$, where $H_S=\frac{m_e^2c^3}{\mid e\mid \hbar}=
4.4\cdot 10^{13}G$ is the Schwinger magnetic field, $v_F=p_F/m_e=(3\pi^2)^{1/3}\hbar n_e^{1/3}/m_e$ is the speed of electrons at the Fermi surface, $\hbar$ is the Planck constant divided by $2\pi$, $m_e$ is the electron rest mass, c is the speed of light in vacuum and $\mid e\mid$ is the magnitude of the electron charge.

In this Rapid Communication, we consider instabilities of the quantized longitudinal electric waves \cite{ltsin} due to an electron beams. In our study, we suppose that the electrons of plasma and beam are degenerate, but not ions. Namely, the equilibrium distribution functions for the plasma and beam electrons are assumed to be the step functions $f_\alpha=H(\mu_\alpha-\varepsilon_\alpha^\ell)$, which equals 1 for $\mu_\alpha=\varepsilon_{F\alpha}\geq\varepsilon_\alpha^\ell=\frac{p_z^2}{2m_\alpha}+2\ell\beta_B H$ (where $\varepsilon_{F\alpha}=\frac{p_{F\alpha}^2}{2m_\alpha}$ is the limiting Fermi energy of the $\alpha$ electrons (e for plasma and b for beam), $p_z$ is the electron momentum in the z-direction, $\beta_B=\frac{\mid e\mid\hbar}{2m_ec}$ is the Bohr magneton, and $\ell$ is the orbital quantum number $\ell=0,1,2,...$), and zero for $\varepsilon_{F\alpha}<\varepsilon_\alpha^\ell$. Note that the kinetic energy of the electrons becomes dependent on the magnetic field due to the quantization of the orbital motion of electrons. We also note here that the maximum of $\ell$ should be $\ell_{max}=1/\eta_\alpha$, since $p_z=p_{F\alpha}(1-\ell\eta_\alpha)^{1/2}$ is real, where $\eta_\alpha=\frac{\hbar\omega_c}{\varepsilon_{F\alpha}}$ and $\omega_c=\frac{\mid e\mid H}{m_ec}$ is the cyclotron frequency of the electron.

In Ref. \cite{tsin}, we have derived a new type of quantum kinetic equations of the Fermi particles. For our purpose, we employ the novel equation with quantum Madelung term
\begin{eqnarray}
\label{tob}
\frac{\partial f_\alpha }{\partial t}+\left( \vec{v}\cdot\nabla \right) f_\alpha
+e_\alpha \Bigl(\vec{E}+\frac{\vec{v} \times \vec{H}}{c}\Bigr)\frac{\partial f_\alpha }{
\partial \vec{p}}+\frac{\hbar^2}{2m_\alpha }\nabla \frac{1}{\sqrt{
n_\alpha }}\Delta \sqrt{n_\alpha }\ \frac{\partial f_\alpha }{\partial \vec{p}}=0 \ ,
\end{eqnarray}
where we have neglected particles collisions and spin.

We linearize Eq.(\ref{tob}) for plasma and beam electrons, the kinetic equation of ions and the Poisson equation with respect to perturbations to obtain
\begin{eqnarray}
\label{pere}
\frac{\partial\delta f_\alpha^\ell}{\partial t}+v\frac{\partial\delta f_\alpha^\ell}{\partial z}+\Bigl(e\frac{\partial\varphi}{\partial z}+
\frac{\hbar^2}{4m_e}\frac{\partial^3}{\partial z^3}\frac{\delta n_\alpha}{n_{0\alpha}}\Bigr)\frac{\partial f_{0\alpha}^\ell}{\partial p_z}=0
\end{eqnarray}
\begin{eqnarray}
\label{peri}
\frac{\partial\delta f_i}{\partial t}+v\frac{\partial\delta f_i}{\partial z}-e\frac{\partial\varphi}{\partial z}
\frac{\partial f_{0i}}{\partial p_z}=0
\end{eqnarray}
\begin{eqnarray}
\label{poi}
\frac{\partial^2\varphi}{\partial z^2}=4\pi e\left\{\sum_\alpha\frac{m_e\eta_\alpha\varepsilon_{F\alpha}}{2\pi^2\hbar^3} 
\sum_{\ell=0}^\infty\int dp_z\delta
f_\alpha^\ell-\int dp_z\delta f_i\right\} \ ,
\end{eqnarray}
where $n_{0\alpha}$ is the equilibrium total number density of $\alpha$ electrons, which as it was shown in Ref.\cite{ltsin} strongly depends on the magnetic field
\begin{eqnarray}
\label{fden}
n_{0\alpha}=\frac{p_{F\alpha}^3}{2\pi^2\hbar^3}\left\{\eta_\alpha+\frac{2}{3}(1-\eta_\alpha)^{3/2}\right\}\ .
\end{eqnarray}
In the above equations (\ref{pere})-(\ref{poi}) the magnetic field is assumed to be directed along the z-axis and the longitudinal wave propagates in the same direction. Moreover, the perturbation of the equilibrium distribution function of the electrons $f_{0\alpha}^\ell$ ( $f_\alpha=\sum_\ell f_\alpha^\ell (z,t,p_z,\ell)$ ) and the ions $f_{0i}$ are small, i.e., $\mid\delta f_\alpha^\ell\mid \ll f_{0\alpha}^\ell$ and $\mid\delta f_i\mid \ll f_{0i}$.

We look for wave solutions in space and time for $\delta f_\alpha^\ell$, $\delta f_i$ and $\varphi$, assuming that they are proportional to $\exp{i(kz-\omega t)}$. We here propose the excitation of longitudinal waves by a straight electron beam with the density $n_{0b}$ much less than the plasma density, which is injected into a degenerate electron gas (ions are assumed to be immobile). To this end, we write the dispersion equation for the plasma - beam system
\begin{eqnarray*}
\label{dde}
1+\delta\varepsilon_e+\delta\varepsilon_b=0
\end{eqnarray*}
\begin{eqnarray}
\label{fdisp}
1-\frac{\omega_{Le}^2}{\Gamma_e}\left\{\frac{1}{\omega^2-k^2v_{Fe}^2}-\frac{2\sqrt{1-\eta_e}}{
\eta_e k^2v_{Fe}^2}\Bigl(1-\frac{\omega}{2kv_{Fe}\sqrt{1-\eta_e}}\ln\frac{\omega+kv_{Fe}\sqrt{1-\eta_e}}{\omega-kv_{Fe}\sqrt{1-\eta_e}}\Bigr)
\right\}-
\nonumber \\
\frac{\omega_{Lb}^2}{\Gamma_b}\left\{\frac{1}{\omega^{\prime 2}-k^2v_{Fb}^2}-\frac{2\sqrt{1-\eta_b}}{
\eta_b k^2v_{Fb}^2}\Bigl(1-\frac{\omega^\prime}{2kv_{Fb}\sqrt{1-\eta_b}}\ln\frac{\omega^\prime+kv_{Fb}\sqrt{1-\eta_b}}{\omega^\prime-
kv_{Fb}\sqrt{1-\eta_b}}\Bigr)
\right\}=0 \ ,
\end{eqnarray}
where
\begin{eqnarray*}
\Gamma_\alpha=1-\Omega_{q\alpha}^2\left\{\frac{1}{\omega^2-k^2v_{F\alpha}^2}-\frac{2\sqrt{1-\eta_\alpha}}{\eta_\alpha k^2v_{F\alpha}^2}\Bigl(1-\frac{\omega }{
2kv_{F\alpha}\sqrt{1-\eta_\alpha}}\ln \frac{\omega +kv_{F\alpha}\sqrt{1-\eta_\alpha}}{\omega -kv_{F\alpha}\sqrt{1-\eta_\alpha}}\Bigr)\right\}\ ,
\end{eqnarray*}
\begin{eqnarray*}
\Omega_{q\alpha}^2=\frac{\hbar^2k^4}{4m_e^2}\ \frac{1}{1+\frac{2}{3\eta_\alpha}(1-\eta_\alpha)^{3/2}}=\frac{\omega_q^2}{1+\frac{2}{3\eta_\alpha}(1-\eta_\alpha)^{3/2}} \ ,
\end{eqnarray*}
$\omega_q$ being the frequency of quantum oscillations of electrons, $\omega^\prime=\omega-\vec{k}\cdot\vec{u}$, $\vec{u}$ is the velocity of beam, and we introduced the Langmuir frequency in the form
\begin{eqnarray}
\label{lfre}
\omega_{L\alpha}^2=\frac{4\pi e^2n_{0\alpha}}{m_\alpha\Bigl(1+\frac{2}{3\eta_\alpha}(1-\eta_\alpha)^{3/2}\Bigr)}=
\frac{\omega_{p\alpha}^2}{1+\frac{2}{3\eta_\alpha}(1-\eta_\alpha)^{3/2}}\ .
\end{eqnarray}
We specifically note here that in the dispersion relation (\ref{fdisp}) the second and fourth terms are due to the quantization, that is in the absence of the magnetic field they don't exist.

We now examine the dispersion equation (\ref{fdisp}), which describes the interaction of low density electron beam with the dense electron plasma, for some interesting cases. First, we assume that $\eta_\alpha >1$ and $\omega^2\gg k^2v_{Fb}^2$, because the beam electron density is less than that of the plasma electrons. In this case, we obtain the dispersion relation
\begin{eqnarray}
\label{disrf}
1-\frac{\omega_{pe}^2}{\omega^2-k^2v_{Fe}^2-\omega_q^2}-\frac{\omega_{pb}^2}{(\omega-\vec{k}\cdot\vec{u})^2-\omega_q^2}=0 \ .
\end{eqnarray}
We emphasize here that in this case the density of both electrons strongly depends on the magnetic field $\omega_{p\alpha}^2=\frac{
p_{F\alpha}^3}{2\pi^2\hbar^3}\ \eta_\alpha$.

If we neglect the last term in Eq.(\ref{disrf}), then we recover the dispersion equation derived in Ref.\cite{ltsin}
\begin{eqnarray}
\label{disrr}
\omega_0^2=\omega_{pe}^2+k^2v_{Fe}^2+\omega_q^2  \ .
\end{eqnarray}
It should be noted that this dispersion relation is quite different than the one obtained by Klimontovich and Silin \cite{kli}, in which it is assumed that $\omega_{pe}^2>k^2v_{Fe}^2$, whereas in Eq.(\ref{disrr}) all terms can be the same order.

We also note that the last term in Eq.(\ref{disrf}) differs from the classical one, where the quantum frequency $\omega_q=0$. In the classical case the interaction of an electron beam with a plasma is strong when the Cherenkov resonance condition ($\omega\simeq\vec{k}\cdot\vec{u}$) is fulfilled. However, in the quantum plasma $\omega_q\neq 0$, and the quantum interaction is governed by first-order poles (the last term in Eq.(\ref{disrf})). Hence, the quantum frequency $\omega_q$ of the de Broglie waves in Eq.(\ref{disrf}) leads to the Doppler resonance alone, $\omega=\vec{k}\cdot\vec{u}-\omega_q$. Equation (\ref{disrf}) therefore admits the solution at $\omega=\omega_0+\gamma$ and $\omega=\vec{k}\cdot\vec{u}-\omega_q+\gamma$, with $\mid\gamma\mid\ll\omega_0$
\begin{eqnarray}
\label{dsol}
Im\gamma=\frac{\omega_{pb}}{2}\ \frac{\omega_{pe}}{\sqrt{\omega_q\omega_0}}\ .
\end{eqnarray}
This expression describes the excitation of new waves $\omega_0=\sqrt{\omega_{pe}^2+k^2v_{Fe}^2+\omega_q^2}$ \cite{ltsin} by the monoenergetic electron beam. It should be noted that the growth rate is purely quantum.

Next in the case when $\hbar\omega_c<\varepsilon_{Fe}=p_{Fe}^2/2m_e$ and $\hbar\omega_c>\varepsilon_{Fb}=p_{Fb}^2/2m_e$, i.e. $\eta_e<1$ and $\eta_b>1$, the dispersion equation (\ref{fdisp}) reduces to
\begin{eqnarray}
\label{rdis}
1-\frac{\omega_{Le}^2}{\Gamma_e}\left\{\frac{1}{\omega^2-k^2v_{Fe}^2}+\frac{2}{3}\frac{(1-\eta_e)^{3/2}}{
\eta_e\omega^2}\right\}-\frac{\omega_{pb}^2}{(\omega-\vec{k}\cdot\vec{u})^2-\omega_q^2}=0 \ .
\end{eqnarray}
Here
\begin{eqnarray}
\label{game}
\Gamma_e=1-\Omega_{qe}^2\left\{\frac{1}{\omega^2-k^2v_{Fe}^2}+\frac{2}{3}\frac{(1-\eta_e)^{3/2}}{
\eta_e\omega^2}\right\}\ .
\end{eqnarray}

In order to solve Eq.(\ref{rdis}), we rewrite it in such form
\begin{eqnarray}
\label{solre}
(\omega^2-\omega_+^2)(\omega^2-\omega_-^2)=\frac{\omega_{pb}^2}{(\omega-\vec{k}\cdot\vec{u})^2-\omega_q^2}
\left\{\omega^2\Bigl(\omega^2-(1+\beta)\Omega_{qe}^2-k^2v_{Fe}^2\Bigr)+\beta\Omega_{qe}^2k^2v_{Fe}^2\right\} \ ,
\end{eqnarray}
where
\begin{eqnarray}
\label{ompl}
\omega_+^2=(1+\beta)(\omega_{Le}^2+\Omega_{qe}^2)+k^2v_{Fe}^2-\omega_-^2 \ ,
\end{eqnarray}
\begin{eqnarray}
\label{ommi}
\omega_-^2=\frac{\beta(\omega_{Le}^2+\Omega_{qe}^2)k^2v_{Fe}^2}{\omega_+^2} \ ,
\end{eqnarray}
$\beta=\frac{2}{3\eta_e}\ (1-\eta_e)^{3/2}$ and $\omega_+^2\gg\omega_-^2$.

Solving Eq.(\ref{solre}) first for the case when $\omega\approx\omega_+$, we assume $\omega=\omega_++\gamma$ and $\omega=\vec{k}\cdot\vec{u}-\omega_q$ to obtain the growth rate after a simple calculation
\begin{eqnarray}
\label{fgr}
Im\gamma=\frac{\omega_{pb}}{2}\ \frac{\omega_{Le}\sqrt{1+\beta}}{\sqrt{\omega_q\omega_+}}\ .
\end{eqnarray}
Next in the case when $\omega\approx\omega_-$, after substitution $\omega=\omega_-+\gamma$ and $\omega=\vec{k}\cdot\vec{u}-\omega_q$, we get the growth rate
\begin{eqnarray}
\label{ngr}
Im\gamma=\frac{\beta^{1/2}}{2(1+\beta)}\ \frac{\omega_{pb}}{\omega_+}\ \frac{k^2v_{Fe}^2}{\sqrt{\omega_-\omega_q}}\ .
\end{eqnarray}
It should be emphasized that the expressions (\ref{ompl}) and (\ref{ommi}) are novel branches of longitudinal waves, and exist only when the orbital motion of electrons in a magnetic field is quantized.

In the range of long wavelengths, when $(\omega_{Le}^2+\Omega_{qe}^2)\gg k^2v_{Fe}^2$, the spectrum $\omega_-$ can be approximated by
\begin{eqnarray}
\label{aommi}
\omega_-^2=\frac{\beta}{1+\beta}k^2v_{Fe}^2 \ .
\end{eqnarray}
We note here that these oscillations of the electron plasma are undamped, and the propagation of these waves is similar to the phonons in Bose medium \cite{lif}.

If we neglect the quantum frequency $\omega_q$ in the beam term, then the growth rate becomes large due to the Cherenkov resonance. This is understandable since a resonance occurs when the beam velocity coincides with the phase velocity of the waves. For the case $\omega=\omega_++\gamma$ and $\omega=\vec{k}\cdot\vec{u}+\gamma$ the growth rate is
\begin{eqnarray}
\label{cfgr}
Im\gamma=\frac{\sqrt{3}}{2}\Bigl(\frac{1+\beta}{2}\ \frac{\omega_{pb}^2\omega_{Le}^2}{\omega_+}\Bigr)^{1/3}\ .
\end{eqnarray}
Whereas for $\omega=\omega_-+\gamma$ and $\omega=\vec{k}\cdot\vec{u}+\gamma$, we get
\begin{eqnarray}
\label{csgr}
Im\gamma=\frac{\sqrt{3}}{2}\ kv_{Fe}
\Bigl(\frac{\beta\omega_{pb}^2kv_{Fe}}{2\omega_-\omega_+^2}\Bigr)^{1/3}\ .
\end{eqnarray}

We now consider the zero sound in a quantized plasma, which is the continuation of the
electron Langmuir waves into the range of short wavelengths, i.e. $k^2r_{TF}^2\gg 1$ (where $r_{TF}$ is the Thomas-Fermi screening length). Oscillations corresponding to the zero sound in a slightly non-ideal Fermi gas was first discussed by Klimontovich and Silin \cite{kli},\cite{lif}, and a dispersion equation of the zero sound in the quantum plasma was recently derived in Ref.\cite{tsin}.

In the following we take into account the quantization effects and obtain an expression for the spectrum of zero sound, and discuss the excitation of this mode. To this end, we assume that $\omega=kv_{Fe}\sqrt{1-\eta_e}+\delta$ (here $\mid\delta\mid\ll kv_{Fe}\sqrt{1-\eta_e}\ $) and $\omega<kv_{Fe}$ in Eq.(\ref{fdisp}), and neglect the beam term. A simple calculation leads to the zero sound spectrum
\begin{eqnarray}
\label{zss}
\omega=kv_{Fe}\sqrt{1-\eta_e}\Bigl(1+2\exp\left\{-\frac{\eta_e}{\sqrt{1-\eta_e}}\ (\frac{k^2v_{Fe}^2}{\omega_{Le}^2+\Omega_{qe}^2}+1+\frac{2\sqrt{1-\eta_e}}{\eta_e})\right\}\Bigr) \ .
\end{eqnarray}
To discuss the excitation of the zero sound by an electron beam, we recall that for $\eta_b>1$ and $\omega=\vec{k}\cdot\vec{u}-\omega_q+\gamma$ ($\ \mid\gamma\mid\ll\omega$), the dielectric permittivity of the beam is
\begin{eqnarray}
\label{bedp}
\delta\varepsilon_b=\frac{\omega_{pb}^2}{2\omega_q\gamma} \ ,
\end{eqnarray}
while in the dielectric permittivity of the plasma $\delta\varepsilon_p$, we suppose 
\begin{eqnarray}
\label{sup}
\omega= kv_{Fe}\sqrt{1-\eta_e}+\gamma \ .
\end{eqnarray}
Substituting Eq.(\ref{bedp}) into Eq.(\ref{fdisp}) and taking into account (\ref{sup}), we get the spectrum (\ref{zss}) for real $\omega$ and for the growth rate
\begin{eqnarray}
\label{zgr}
Im\gamma=\Bigl(\frac{\eta_e}{\sqrt{1-\eta_e}}\Bigr)^{1/2}\ \frac{kv_{Fe}}{\omega_q}\ \frac{\omega_{pb}\omega_{Le}kv_{Fe}}{\omega_{Le}^2+\Omega_{qe}^2}\ e^{-\frac{\eta_e}{\sqrt{1-\eta_e}}\ \Bigl(\frac{k^2v_{Fe}^2}{\omega_{Le}^2+\Omega_{qe}^2}+1+\frac{2\sqrt{1-\eta_e}}{\eta_e}\Bigr)}\ .
\end{eqnarray}

We specifically note here that in the expression (\ref{zss}) the second term should be smaller than one. That is the exponent must be rather small, for which the necessary condition without the quantization, as mentioned above, is $ k^2v_{Fe}^2\gg\omega_{pe}^2$. It is clear from Eq.(\ref{zss}) that the quantization of the energy of electrons imposes a new condition. Namely, the zero sound can exist even at  $k^2v_{Fe}^2\simeq\omega_{pe}^2$. In which case the growth rate (\ref{zgr}) tends to zero as $\eta_e\rightarrow 1$.

We next consider the excitation of Bogolyubov's type of spectrum, derived in Ref.\cite{tsin}, by a strong electric field. For this purpose, we replace $\omega$ by $\omega-\vec{k}\cdot\vec{u}$ in the electrons part of the dielectric permittivity, and assume ions to be cold, i.e. $\delta\varepsilon_i=-\frac{\omega_{pi}^2}{\omega^2}$. The dispersion equation for the electron - ion plasma then reads
\begin{eqnarray}
\label{bdde}
1+\delta\varepsilon_e(\omega-\vec{k}\cdot\vec{u}, \vec{k})+\delta\varepsilon_i=0\ .
\end{eqnarray}
In the range of frequencies $\omega\gg kv_{Fi}$ and $\mid\omega-\vec{k}\cdot\vec{u}\mid\ll kv_{Fe}\sqrt{1-\eta_e}$, we obtain
\begin{eqnarray}
\label{dp}
\delta\varepsilon_e=\frac{\omega_{Le}^2\left\{1+\imath\pi\frac{\omega-\vec{k}\cdot\vec{u}}{
2kv_{Fe}\sqrt{1-\eta_e}}\right\}}{k^2v_{Fe}^2+\Omega_{qe}^2\left\{1+\frac{2\sqrt{1-\eta_e}}{\eta_e}\Bigl(
1+\imath\pi\frac{\omega-\vec{k}\cdot\vec{u}}{
2kv_{Fe}\sqrt{1-\eta_e}}\Bigr)\right\}}\ .
\end{eqnarray}

If we assume that the quasi neutrality $n_e+n_i=0$ is satisfied, we can then neglect one in Eq.(\ref{bdde}). With this assumption Eq.(\ref{bdde}) admits the complex roots ($\omega=\omega^\prime+i\omega^{\prime\prime}$) such as
\begin{eqnarray}
\label{comp1}
\omega^{\prime 2}=\frac{m_e}{m_i}\Bigl(\frac{\eta_e+\frac{2}{3}(1-\eta_e)^{3/2}}{\eta_e+2\sqrt{1-\eta_e}}\
k^2v_{Fe}^2+\omega_q^2\Bigr)
\end{eqnarray}
and the imaginary part
\begin{eqnarray}
\label{comp2}
\omega^{\prime\prime}=-\frac{\pi}{2}\frac{kp_{Fe}}{m_i}\ \frac{\eta_e+\frac{2}{3}(1-\eta_e)^{3/2}}{(\eta_e+2\sqrt{1-\eta_e})^2}\ \Bigl(1-\frac{
\vec{k}\cdot\vec{u}}{\omega^\prime}\Bigr)\ .
\end{eqnarray}
For an instability the following inequality should be satisfied
\begin{eqnarray}
\label{sin}
u>\sqrt{\frac{\eta_e+\frac{2}{3}(1-\eta_e)^{3/2}}{\eta_e+2\sqrt{1-\eta_e}}\ \frac{p_{Fe}^2}{m_e m_i}+\frac{\hbar^2k^2}{4m_e m_i}} \ .
\end{eqnarray}
We note here that in the absence of the external magnetic field $\eta_e=0$ the expressions (\ref{comp1}) and (\ref{comp2}) reduce to the expressions of real $\omega$ and damping rate obtained in Refs.\cite{tsin}, \cite{tsin10}.

To summarize, we have studied the excitation of longitudinal waves by an electron beam in a magnetized quantum plasma, assuming that the electrons of the plasma and beam are degenerate, but not ions. We have investigated instabilities of the quantized longitudinal electric waves by a newly derived dispersion equation disclosing a novel branches. We obtained the growth rates of these new modes. We specifically not here that these waves exist only when the orbital motion of electrons in a magnetic field is quantized. We also note here that these oscillations of the degenerate Fermi plasma are undamped, and the propagation of these waves is similar to the phonons in Bose medium. The generation of zero sound by an electron beam is also considered. It is shown that the quantization of the energy of electrons imposes a new condition for the excitation of the zero sound. Furthermore, we discuss the breed of Bogolyubov's type of spectrum by a strong electric field. These investigations may play an essential role for the description of complex phenomena that appear in dense astrophysical objects, as well as in the next generation intense laser-solid density plasma experiments.

\end{document}